\documentclass[
aps,%
12pt,%
final,%
notitlepage,%
oneside,%
onecolumn,%
nobibnotes,%
nofootinbib,
superscriptaddress,%
noshowpacs,%
centertags]%
{revtex4}
\RequirePackage{graphicx}

\def\refitem#1{\relax}

\begin{document}
\title{Identity method - a new tool for studying chemical fluctuations}

\author{\firstname{M.} \surname{Ma\'{c}kowiak} for the NA49 Collaboration}
\email{majam@if.pw.edu.pl}
\affiliation{Faculty of Physics, Warsaw University of Technology, Poland}

\begin{abstract}
Event-by-event fluctuations of the chemical composition of the
hadronic system produced in nuclear collisions are believed to be
sensitive to properties of the transition between confined
and deconfined strongly interacting matter.

In this paper a new technique for the study of chemical fluctuation,
the identity method, is introduced and its features are discussed.
The method is tested using data on
central Pb-Pb collisions at 40$A$~GeV registered by the NA49
experiment at the CERN SPS.
\end{abstract}

\maketitle

\section{Introduction}
The most interesting features of the phase diagram of 
strongly interacting matter are the  
Critical Point (CP) and the $1^{st}$ order phase transition line. 
Event-by-event fluctuations of the chemical 
composition of the hadronic system produced in nuclear collisions are 
believed to be sensitive to both of them. 
The first relevant measurements
were performed by
the NA49 experiment at the CERN SPS. A systematic scan in beam 
energy and system size was recently started by 
the NA61 collaboration. Furthermore, additional insight
is expected from the RHIC beam energy scan program. 

There are  several measures used to quantify chemical fluctuations,
among them: $\sigma_{dyn}$  
\cite{sigma_dyn,kpi_fluct_na49, dima_cpod2009} used by  NA49 
and $\nu_{dyn}$ \cite{nu_dyn_STAR} used by 
STAR. 
Both are related as $\sigma_{dyn}^2\approx\nu_{dyn}$ 
and share the same disadvantage, namely, they depend
on volume (number of wounded nucleons) and volume
fluctuations
in thermodynamical (wounded nucleon) models.  
Another measure $\Phi_{x}$~\cite{phi, Marek} was used by NA49  to 
characterize transverse momentum \cite{phipt_syssize, phipt_energy}, 
electric charge \cite{delta_q} and 
azimuthal angle fluctuations \cite{phiphi}. 
It is free of the mentioned disadvantage of $\sigma_{dyn}$, $\nu_{dyn}$.
However,
all these measures of chemical fluctuations are affected by non-perfect
particle identification. 
This is illustrated in Fig. 1 (left panel), where the spectrum of the specific 
energy loss (dE/dx) of particles  measured by the NA49 Time Projection
Chambers~\cite{nim} is shown for a selected phase-space bin.
The dE/dx signal depends on the particle mass and together with
the particle charge measurement is used to identify particles.  
It is seen that the dE/dx distributions of different
particle species partly overlap and thus unique
particle identification is not possible.

The identity method adapts the $\Phi_{x}$ measure to take into account
a non-unique particle identification, while  
keeping the advantages of $\Phi_{x}$.

\section{Identity method}

Let us assume that
particles are identified according to their measured mass. 
The measured mass spectra  of all particles and of
particles of type $h$ in the analyzed event sample are denoted as 
$\rho$ and $\rho_{h}$, respectively. 
The spectra are normalized to the corresponding  mean multiplicities per event, namely:\newline
\begin{equation}
\begin{tabular*}{10cm}{@{\extracolsep{\fill}} l r }
 $\int dm \, \rho(m) = \langle N \rangle $~, & $\int dm \, \rho_{h}(m) 
= \langle N_{h} \rangle$~. \\
\end{tabular*}
\end{equation}
Furthermore, we
define a single particle variable called the particle identity as:
\begin{equation}
w_{h}(m) \buildrel \rm def \over = \frac{\rho_{h}(m)}{\rho(m)}~.
\end{equation}
The fluctuation measure $\Psi_{wh}$ is then introduced in a way similar (the roots over
the two components are absent) to the
$\Phi$ measure. First, a single 
particle variable $z$ is defined as:
\begin{equation}
z \buildrel \rm def \over = w_{h} - \overline{w_{h}}~,
\end{equation}
where the bar denotes the inclusive mean and thus 
$
\overline{w} = \langle N_{h} \rangle / \langle N \rangle~.
$
Second,  an event variable $Z$, which is the multiparticle analog of $z$, 
is calculated as:
\begin{equation}
Z \buildrel \rm def \over = \sum_{i=1}^{N}(w_{h}(m_{i}) - \overline{w_{h}})~,
\end{equation}
where $ N $ is the multiplicity and $ i $ is the particle index in an event.\newline
Finally, the fluctuation measure $\Psi_{wh}$ is defined as:
\begin{equation}
\label{Psi-def}
\Psi_{wh} \buildrel \rm def \over = 
\frac{\langle Z^2 \rangle}{\langle N \rangle} - \overline{z^2}~.
\end{equation}
For further analysis one denotes two possible values of $\Psi_{wh}$:
\begin{itemize}
\item $\Psi_{res}$ which is the value of $\Psi_{wh}$ for the experimental mass resolution case,
\item $\Psi_{corr}$ which is the value of $\Psi_{wh}$ for the perfect mass resolution case.
\end{itemize}
In order to correct for the non-unique particle identification we calculate
the variance per particle due to random identification for the 
experimental mass resolution case:
\begin{equation}
\label{Var-def}
Var_{res} = \frac{1}{\langle N \rangle} 
\int_{0}^{\infty} dm \, \rho(m) \cdot w_{h}(m)(1-w_{h}(m))~.
\end{equation}
It is easy to show that for unique particle identification
(the perfect mass resolution case, $w_{h}(m) = \delta(m - m_h)$) 
the result is $Var_{res} \buildrel \rm def \over = Var_{A} = 0$, 
whereas for no mass resolution ( $w_{h}(m) = const.$ ) one obtains
$Var_{res} \buildrel \rm def \over = Var_{B} = \frac{\langle N_{h} \rangle}{\langle N \rangle} (1-\langle N_{h} \rangle /
\langle N \rangle )$. 
For the experimental data analysis the  integral in Eq.~\ref{Var-def}
 is replaced by a sum over all particles. \newline
The following key relation can be proven 
\footnote{M. Ga\'{z}dzicki, K. Grebieszkow, M. Ma\'{c}kowiak and S. Mr\'{o}wczy\'{n}ski 
publication in preparation}:
\begin{equation}\label{Id-def}
\Psi_{corr} = \Psi_{res}(1-Var_{res} / Var_{B})^{-2}~.
\end{equation}
It shows that the measured fluctuations $\Psi_{res}$ can be
corrected for the effect of non-unique particle identification
in a model independent way. This is because the correction factor,
$(1 - Var_{res}/ Var_{B})^{-2}$, depends only on the experimental resolution
and mean particle multiplicities.

\begin{figure}[h]
\begin{minipage}{15pc}
\vspace{-1pc}
\hspace{-4pc}
\includegraphics[width=15pc]{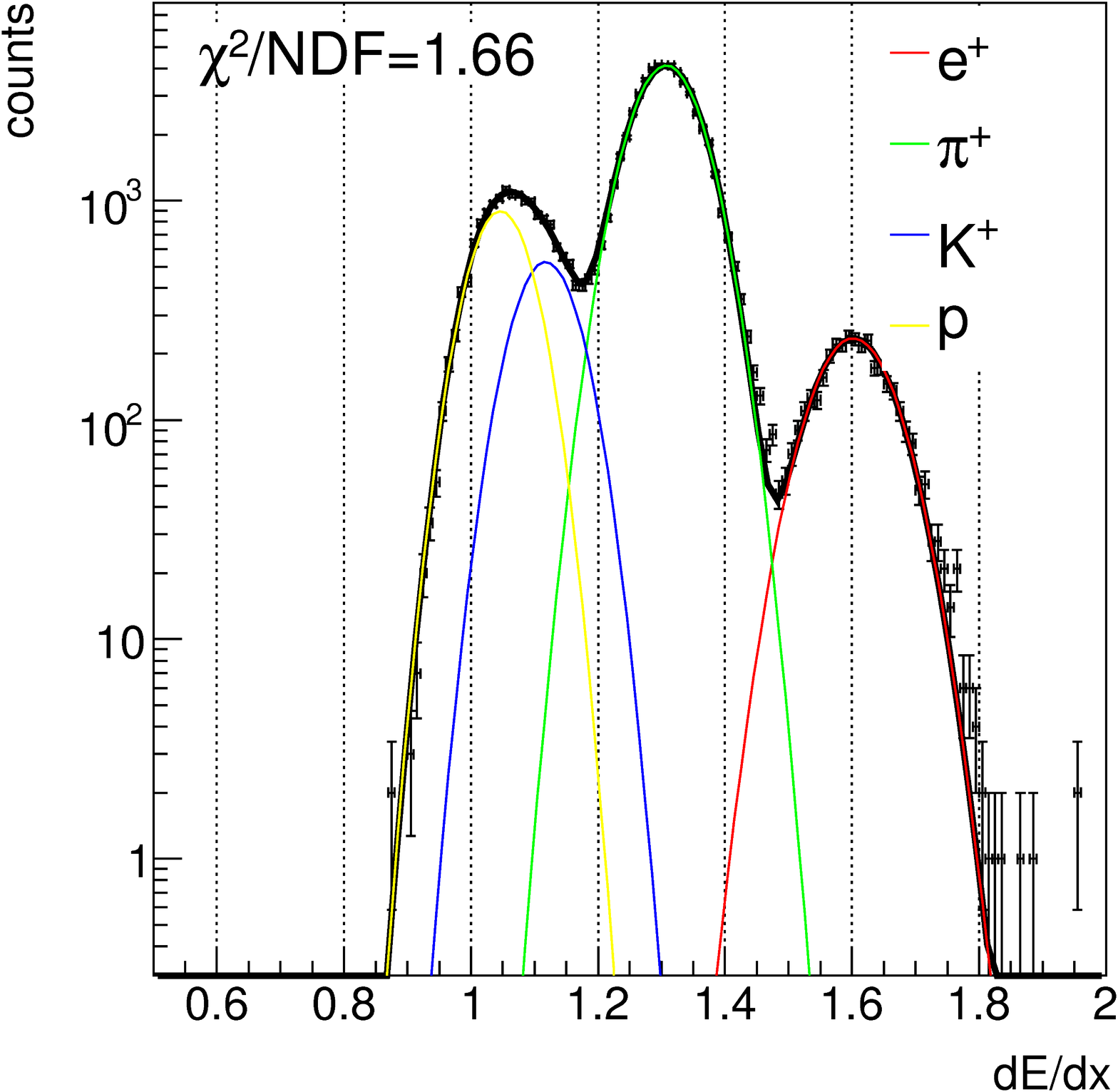}
\end{minipage}\hspace{1pc}%
\begin{minipage}{15pc}
\includegraphics[width=17pc]{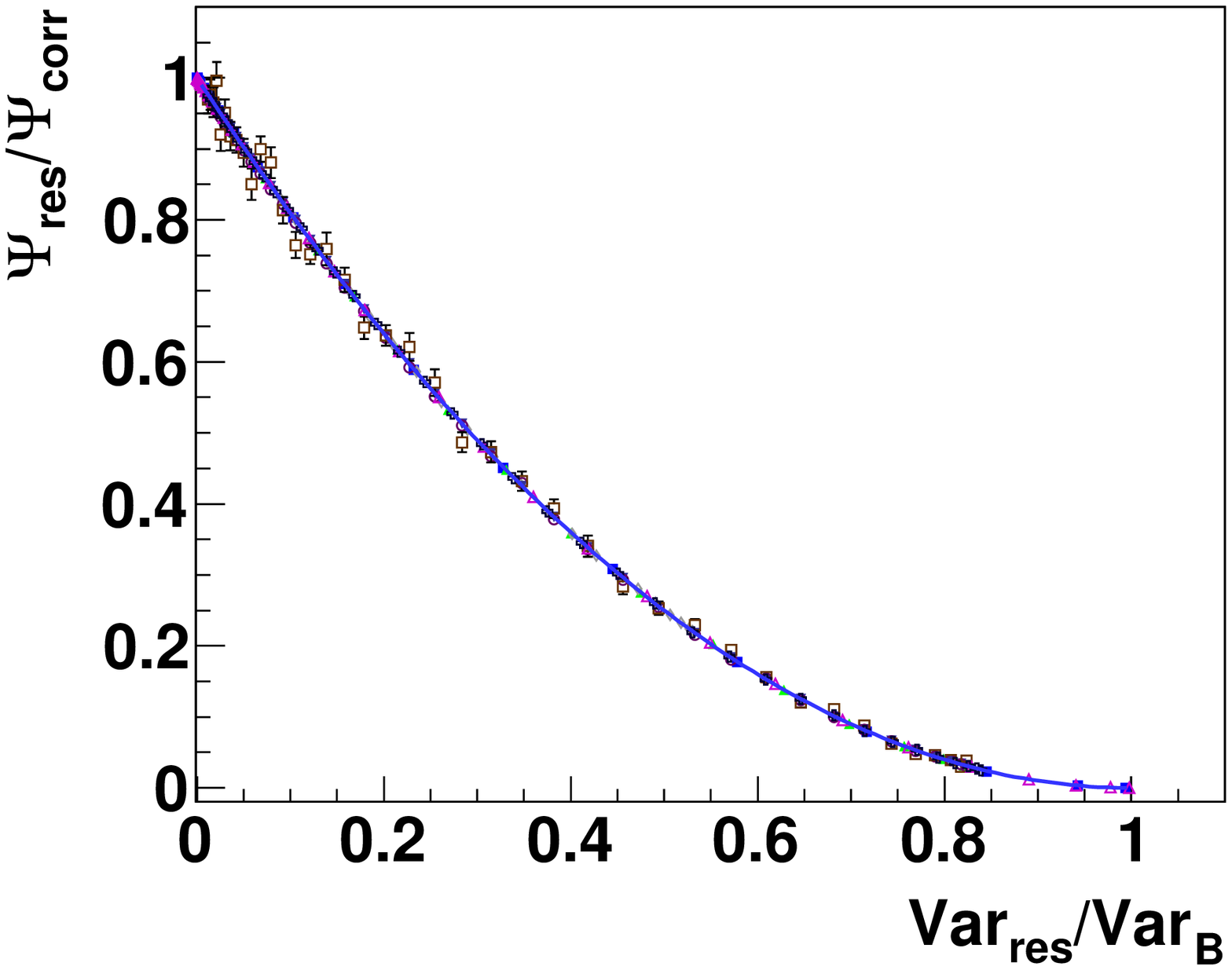}
\end{minipage}\hspace{1pc}%
\caption{
Left panel:  
Distribution of specific energy loss measured in the 
NA49 TPCs for positively charged particles is a bin
$p_{tot}\epsilon(4.4-5.3)$~GeV/c, 
$p_{T}\epsilon(0.0-0.2)$~GeV/c and 
$\phi\epsilon(0.75\pi-\pi)$.
The fitted Gauss functions are shown by solid curves.
Right panel: 
The ratio $\Psi_{res}/\Psi_{corr}$ versus  $Var_{res}/Var_B$ calculated within several
Monte Carlo simulations with different parameters of
experimental resolution, particle multiplicities and
fluctuations.
The results agree with the analytical dependence given by Eq.~\ref{Id-def}.
}
\label{Psi}
\end{figure}

Equation~\ref{Id-def} was checked by numerous  Monte Carlo simulations 
with different types of correlations,
mass resolution functions and mean particle multiplicities. 
The results of these simulations are shown in 
Fig.~\ref{Psi} (right panel).

\section{The identity method test using NA49 data}

In the analysis of experimental data we use the particle energy loss $dE/dx$ as the measure of the
mass $m$.
For optimal identification the dE/dx spectra from the NA49 TPCs are determined in bins of 
total and transverse momentum, azimuthal angle as well as for
both electric charges separately~\cite{sigma_dyn,kpi_fluct_na49,dima_cpod2009}. 
In each  bin four 
Gauss functions (for electrons, pions, kaons and protons) are fitted. 
An example of such a fit is displayed in 
Fig.~1 (left panel). 
The fitted functions are then used as the $\rho_{h}$ and $\rho$ functions 
of the identity method.
The further analysis steps are as follows:
\begin{itemize}

\item using mean particle multiplicities the variance  
      $Var_{B} = \langle N_{h} \rangle / \langle N \rangle 
      ( 1 - \langle N_{h} \rangle / \langle N \rangle )$ 
      is obtained, 

\item for each particle its  identity   is calculated
      \begin{equation}
      w_{hi}(<dE/dx>_{i}|q,p_{tot},p_{T},\phi) 
      = \frac{\rho_{h}(<dE/dx>_{i}|q,p_{tot},p_{T},\phi)}
       {\rho(<dE/dx>_{i}|q,p_{tot},p_{T},\phi)}~,
       \end{equation}

\item using the experimental dE/dx resolution functions,
      $\rho_{h}$ and $\rho$ (M is the total number of particles used in the analysis) the variance  
      \begin{equation}
      Var_{res} = \frac{1}{M} \sum_{1}^{M} w_{hi}
      (<dE/dx>_{i}|q,p_{tot},p_{T},\phi)(1-w_{hi}
      (<dE/dx>_{i}|q,p_{tot},p_{T},\phi))~,
      \end{equation}
      is computed,

\item using the identity values, $w_{hi}$, $\Psi_{res}$ is calculated and
 
\item the corrected value of $\Psi_{corr}$ is obtained using Eq.~\ref{Id-def}.

\end{itemize}

As a first test of the identity method proton fluctuations 
were studied in Pb+Pb collisions at 40$A$~GeV energy.
Positively and negatively charged particles
with total momentum 
up to 40 GeV/c and transverse momentum up to 2 GeV/c 
were used for the analysis. 
The total number of analyzed events was 4000. 
The mean multiplicities are $<N>=165.40$, and $<N_{p}>=42.16$. 
The obtained  
value of $\Psi_{res} \cdot 1000 = -17.4 \pm 3.5$. 
The corresponding correction factor for non-unique
particle identification was calculated to be $\approx 1.2$.\newline
The value corrected for the finite resolution is $\Psi_{corr} \cdot 1000 = -22.3 \pm 4.4$. 
The analysis of Pb+Pb 
collisions at all NA49 energies is in progress.

\newpage

\newpage

\begin{center}
FIGURE CAPTIONS
\end{center}
\begin{enumerate}
\item
Left panel:  
Distribution of specific energy loss measured in the 
NA49 TPCs for positively charged particles is a bin
$p_{tot}\epsilon(4.4-5.3)$~GeV/c, 
$p_{T}\epsilon(0.0-0.2)$~GeV/c and 
$\phi\epsilon(0.75\pi-\pi)$.
The fitted Gauss functions are shown by solid curves.
Right panel: 
The ratio $\Psi_{res}/\Psi_{corr}$ versus  $Var_{res}/Var_B$ calculated within several
Monte Carlo simulations with different parameters of
experimental resolution, particle multiplicities and
fluctuations.
The results agree with the analytical dependence given by Eq.~\ref{Id-def}.
\end{enumerate}

\end{document}